\documentclass[aps,showpacs,pra,twocolumn]{revtex4}
\usepackage{mathrsfs} 
\usepackage{amssymb}

\usepackage{graphicx}
\usepackage{dcolumn}
\usepackage{bm}
\usepackage{mathtools}

\begin{document}


\title{Quantum Random Walks of Waves}



\author{Tian-Li Feng, Yong-Sheng Zhang$\footnote{
Email: yshzhang@ustc.edu.cn}$, Guang-Ming Zhao, Sheng Liu, and Guang-Can Guo} \affiliation{Key Laboratory of Quantum Information, University of Science and Technology of
China, CAS, Hefei, 230026, People's Republic of China}
\date{\today }

\pacs{05.40.Fb, 03.65.Ta, 03.67.-a}




\begin{abstract}
  %
  The extremely fascinating behaviors of the quantum walks of particles, which differ much from the classical counterparts, have attracted many attentions. Here we investigate another interesting part of the quantum walks, that is the quantum walks of waves. Firstly, we show the behaviors of the quantum walks of plane wave, which are largely different from the counterparts of either the classical or the quantum walks of particles. Two situations -- with and without intermediate time measurements of the walks are considered. At last, it is shown that the quantum walks of plane wave can be used to calculate the evolution of the general wave packets, e.g., Gaussian wave packet.

\end{abstract}


\maketitle

\section{Introduction}
  In recent years quantum walks have gained great interest from physicists, mathematicians, computer scientists, and engineers. This is caused by the fascinating prospect of the applications of the quantum walks to the algorithm in the quantum computer, which is supposed to be born in the near future. Recent demonstration of a search algorithm \cite{quantum algorithm,Shenvi} based on quantum random walks showed that it is exponentially faster to solve a problem by the quantum walks than by the best classical algorithm.

  The behavior of quantum walks is extremely different from the classical random walks. As same as the classical case, the quantum walks also have a coin(or many coins) and a particle. The particle moves in one direction based on the result of the tossed coin. However, in quantum case, there can be no measurement of the coin states during the process of the walks, moreover, both the flip of the coin and the conditional motion of the particle are unitary transformations, so the process is reversible. The variance of the probability of the particle's position, which grows linearly with time in the classical walk, by contrast, grows quadratically with time in the quantum walk, for the reason of the interference between the possible paths of the particle in the quantum walks when there is no measurement at intermediate time steps.

  The quantum walks of particles have been studied very well in recent years. Since the quantum random walks has been proposed by Aharonov \emph{et al.} \cite{Aharonov1}, many aspects of the quantum walks have been much studied. Nayak and Vishwanath have got the exact solutions of the Hadamard walk of particles in one dimension in Ref.\,\cite{Nayak}. Quantum walks include one dimension \cite{Aharonov1,Ambainis,Konno1,Konno2,T.A.Brun1,T.A.Brun2,T.A.Brun3,Nayak,Yutaka Shikano} and high dimensions in graphs \cite{Mackay,Aharonov2,complementarity}, discrete and continuous \cite{Continuous and discrete}, periodic and quasi-periodic or random \cite{Ribeiro}, etc. T. Brun \emph{et al.} have studied much on the multi-coin and decoherence walks \cite{T.A.Brun1,T.A.Brun2,T.A.Brun3,Zhang Kai}, and the absorbing problems have been well discussed in \cite{Konno2,Yamasaki}.

  However, the quantum walks of waves have not been much studied yet. Using the long-wave approximation, Aharonov \emph{et al.} have studied the random walks of a Gaussian wave packet in Ref.\,\cite{Aharonov1}, in which the walk is set to be measured after each step. We will review this work in the following sections.

  Here, we proceed as follows: first we give a brief review of the quantum walks of particles in section II, in which we also do some improvement of the walks. On the one hand, this section would acts as a contrast with the quantum walks of waves described in the following sections; on the other hand, it would be more convenient to describe the walks of waves by introducing the walks of particles first. In section III we would turn to the study of quantum walks of waves, which consists of the brief introduction to the background of the quantum walks of waves proposed in Ref.\,\cite{Aharonov1} and the study of quantum random walks of plane wave, including both the cases with measurements and without any measurement in the intermediate time steps. At last, we will use the results of walks of plane wave to study the walks of general wave packets in section IV, in which we will give some numerical simulation of the result and compare some of them with Ref.\,\cite{Aharonov1}.

\section{The quantum walks of particles}
  In the discrete quantum walks, each step consists of two parts: the flip of coin and the condition motion of the particle. The value of the coin controls the motion of the particle, with the state $|R\rangle$ to go the right direction by a distance of $l$, and the state $|L\rangle$ means to go left. Generally, quantum walks are directed by the unitary evolution operator.

  For an initial state:
  \begin{equation}
  |\Phi_0\rangle=\sum_x[a_{\scriptscriptstyle R}(x,0)|R\rangle+a_{\scriptscriptstyle L}(x,0)|L\rangle]\otimes|x\rangle,
  \end{equation}
  the unitary evolution operator can be described as
  \begin{equation}
  \hat{E}=(\hat{P}_R\otimes\hat{S}+\hat{P}_L\otimes\hat{S}^\dagger)(\hat{C}\otimes\hat{I}).
  \end{equation}
  Here the general unitary coin operator
  \begin{equation}
  \hat{C}=\frac{e^{i{\eta}}}{\sqrt{2}}\left(\begin{array}{cc}e^{i{\phi}} & 0  \\0 & e^{-i{\phi}} \end{array}\right)\left(\begin{array}{cc}e^{i{\theta}} & e^{-i{\theta}}  \\e^{-i{\theta}} & -e^{i{\theta}} \end{array}\right)\left(\begin{array}{cc}e^{i{\varphi}} & 0  \\0 & e^{-i{\varphi}} \end{array}\right).
  \end{equation}
  The space operator is
  \begin{equation}
  \hat{U}=\hat{P}_R\otimes\hat{S}+\hat{P}_L\otimes\hat{S}^\dagger,
  \end{equation}
  with
  \begin{equation}
  \hat{S}=e^{-i\hat{P}l/\hbar},\ \ \ \hat{S}^\dagger=e^{i\hat{P}l/\hbar},
  \end{equation}
  $$\hat{S}|x\rangle=|x+l\rangle,\ \ \ \hat{S}^\dagger|x\rangle=|x-l\rangle,$$
  $$\hat{P}_R=|R\rangle\langle R|, \ \ \ \hat{P}_L=|L\rangle\langle L|.$$

  After $t$ steps, without any measurements during the process of the walks, the state of the particle becomes
  $$|\Phi(t)\rangle=\hat{E}^t|\Phi(0)\rangle=\sum_x[a_{\scriptscriptstyle R}(x,t)|R\rangle+a_{\scriptscriptstyle L}(x,t)|L\rangle]\otimes|x\rangle.$$

  The general case with three variable parameters $\eta$, $\phi$, and $\theta$ has been studied in the Ref.\,\cite{Konno1}. One special case, with $\phi=\eta=\theta=0, l=1$, which is the well known Hadamard walk $H=\frac{1}{\sqrt{2}}\left(\begin{array}{cc}1 & 1  \\1 & -1 \end{array}\right)$, has been much studied \cite{T.A.Brun1,Nayak}. By either the way of Fourier analysis or the way of combinatorial analysis one can get the following analytical result: (in Ref.\,\cite{T.A.Brun1})


\begin{equation}
|0\rangle\otimes|R\rangle\xrightarrow[H]{t\ steps}a_{\scriptscriptstyle R}(x,t)|R\rangle+a_{\scriptscriptstyle L}(x,t)|L\rangle,
\end{equation}
\begin{equation}
|0\rangle\otimes|L\rangle\xrightarrow[H]{t\ steps}b_{\scriptscriptstyle R}(x,t)|R\rangle+b_{\scriptscriptstyle L}(x,t)|L\rangle.
\end{equation}

  However, if we reverse the sequence of the space operator and coin operator, that is, we can carry the condition motion first and then use the coin flip at each step, another very different result will be obtained.

 This process may be useful in some cases, for example, the measurement of a component of the spin of a spin-1/2 particle. One particle moves conditionally first in the electromagnetic field by the different force caused by the different spin states in the electromagnetic field, and then make the spin transformation which is correspondent to the coin flip. For the general case of $\phi,\eta,\theta$, one can get the result in the same way.

\section{The Quantum walks of cosine wave}

  We start from the introduction of the work Aharonov \emph{et al.} have done in Ref.\,\cite{Aharonov1}, which is mathematically equivalent to the process introduced in Ref.\,\cite{Kempe}. There is a wave packet with the coin state $|\Psi_{\rm in}\rangle$ and initial space state $f(x,0)$. The walk of the wave packet at each step consists of two parts: letting the operators $\hat{U}$ and $R(\theta)$ act on its state sequentially, and measuring both coin and space states and then re-initializing the coin state into $|\Psi_{\rm in}\rangle$.

\begin{equation}
|\Psi_{\rm in}\rangle=a_{\scriptscriptstyle R}|R\rangle+a_{\scriptscriptstyle L}|L\rangle,
\end{equation}
\begin{equation}
\hat{U}=|R\rangle\langle R|\otimes e^{-i\hat{P}l/\hbar} + |L\rangle\langle L|\otimes e^{i\hat{P}l/\hbar},
\end{equation}
\begin{equation}
R(\theta)=\left(\begin{array}{cc}{\cos\theta} & {-\sin \theta}  \\{\sin\theta} & {\cos\theta} \end{array}\right).
\end{equation}

  However, to get a good result of the walks, one must use the long-wave approximation, that is, the width of the wave is much larger than the step length $l$. By this approximation, the wave packet will just be translated in space with no variance of shape each step. What's more, the moving distance in one step can be much larger than the step length $l$ with a low probability if we choose some special values of $\theta$. The results of Gaussian wave without long-wave approximation can be simulated in numerical method.

  Now we are interested in two things. First, for a general wave packet, what the results will be without any approximation. Second, what the wave packet will be after $t$ steps without measurement in the process of walking. However, the direct calculation is very complicated and difficult, so we will study the long-wave limit---the plane wave first, and then expand the results to the general wave packets. (In contrast, the short-wave limit, which is actually particles, has been reviewed in section II.)

  Before we study the quantum walks of plane waves, it is necessary to note the fact that two plane waves with the same frequency can be combined into one plane wave with no change in frequency, i.e.,

\begin{equation}
Ae^{ik(x+a)}+Be^{ik(x+b)}=Ce^{ik(x+c)},
\end{equation}
with
\begin{equation}
C=\sqrt{A^2+B^2+2AB\cos k(a-b)},
\end{equation}
\begin{equation}
\tan kc=\tan {\frac{A\sin ka+B\sin kb}{A\cos ka+B\cos kb}}\equiv\tan\frac{\alpha}{\beta},
\end{equation}
\begin{equation}
kc=\left\{\begin{array}{c} \arctan(\alpha/\beta)\ \ \ \ \ \ \ \ \ \ \beta>0 \ \ \ \ \ \ \ \ \ \ \\ \pi+\arctan(\alpha/\beta)\ \ \ \ \beta<0,\alpha>0 \\ \arctan(\alpha/\beta)-\pi\ \ \ \ \beta<0,\alpha<0 \end{array}\right.\ .
\end{equation}

\subsection{With measurement after each step}

  It is assumed that the initial state of the plane wave is $|\Psi_{\rm in}\rangle\otimes |e^{ik(x-l_0)}\rangle$, then the operate $\hat{U}$ and operator $\hat{R}(\theta)$ are carried on it sequentially, the coin state will be measured and re-initialized into $|\Psi_{\rm in}\rangle$. After repeating this work several times the state will become $|\Psi_{\rm in}\rangle\otimes |e^{ik(x-l')}\rangle$. Now we consider the next step of the wave:
\begin{eqnarray*}
|\Psi'\rangle &=& \hat{U}|\Psi_{\rm in}\rangle\otimes |e^{ik(x-l')}\rangle \\
&=& a_{\scriptscriptstyle R}e^{ik(x-l'-l)}|R\rangle+a_{\scriptscriptstyle L}e^{ik(x-l'+l)}|L\rangle ,
\end{eqnarray*}
\begin{eqnarray}
\hat{R}(\theta)\hat{S}|\Psi'\rangle &=&
\left(\begin{array}{cc}{\cos\theta} & {-\sin \theta}  \\{\sin\theta} & {\cos\theta} \end{array}\right)\left(\begin{array}{c} a_{\scriptscriptstyle R}e^{ik(x-l'-l)} \\ a_{\scriptscriptstyle L}e^{ik(x-l'+l)} \end{array}\right) \nonumber \\
&=&\left(\begin{array}{c} \sqrt{p_{\scriptscriptstyle R}}e^{ik(x-l'+l_1)} \\ \sqrt{p_{\scriptscriptstyle L}}e^{ik(x-l'+l_2)} \end{array}\right),
\end{eqnarray}
with
\begin{equation}
p_{\scriptscriptstyle R}=|a_{\scriptscriptstyle R}\cos \theta|^2+|a_{\scriptscriptstyle L}\sin \theta|^2-2a_{\scriptscriptstyle R}a_{\scriptscriptstyle L}\sin \theta \cos \theta \cos 2kl ,
\end{equation}
\begin{equation}
p_{\scriptscriptstyle L}=|a_{\scriptscriptstyle R}\sin \theta|^2+|a_{\scriptscriptstyle L}\cos \theta|^2+2a_{\scriptscriptstyle R}a_{\scriptscriptstyle L}\sin \theta \cos \theta \cos 2kl ,
\end{equation}
\begin{equation}
\tan k l_1=\frac{-a_{\scriptscriptstyle R}\cos \theta-a_{\scriptscriptstyle L}\sin \theta}{a_{\scriptscriptstyle R}\cos \theta-a_{\scriptscriptstyle L}\sin \theta} \tan kl,
\end{equation}
\begin{equation}
\tan k l_2=\frac{-a_{\scriptscriptstyle R}\sin \theta+a_{\scriptscriptstyle L}\cos \theta}{a_{\scriptscriptstyle R}\sin \theta+a_{\scriptscriptstyle L}\cos \theta} \tan kl.
\end{equation}
  We can get an amazing result: from the equations above we can see that the walking probability to the left or the right and the walking length are independent of the steps $t$. That is to say, in the whole process the frequency of the wave does not change, and in each step the plane wave moves to right or left with a fixed probability, and the moving distance is fixed too. This is similar to the result achieved in Ref.\,\cite{Aharonov1}, but there is no approximation here.

  After $t$ steps, the average moving distance of the phase of the plane wave is: $\langle x\rangle=t(p_{\scriptscriptstyle R}l_1+p_{\scriptscriptstyle L}l_2)$ , the variance $\sigma^2=tp_{\scriptscriptstyle R}p_{\scriptscriptstyle L}(l_1-l_2)^2$. By the way, the value of $x+2n\pi/k$ has the same meaning with the value of $x$ for plane wave.
\begin{figure}[tbph]
\centering
\includegraphics[width= 3in]{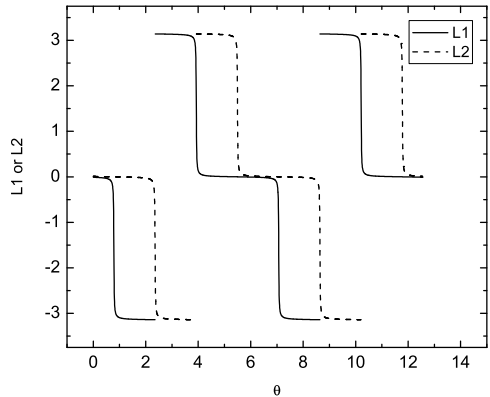}
\caption{The values of $L_1$ and $L_2$ act as the function of $\theta$. With $k=1,\ l=0.01,\ a_{\scriptscriptstyle R}=a_{\scriptscriptstyle L}=1/\sqrt{2}$.}\label{fig1}
\end{figure}
\begin{figure}[tbph]
\centering
\includegraphics[width= 3in]{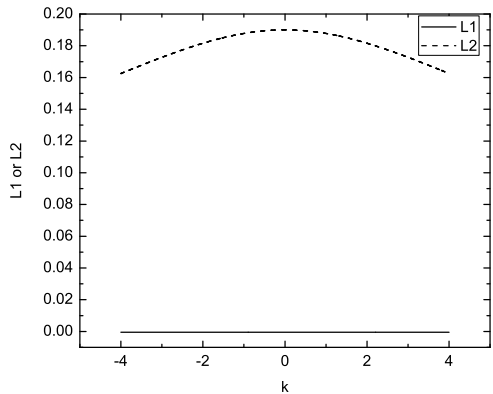}
\caption{The values of $L_1$ and $L_2$ act as the function of frequency $k$. With $l=\theta=1$, $a_{\scriptscriptstyle R}=a_{\scriptscriptstyle L}=1/\sqrt{2}$.}\label{fig2}
\end{figure}

Now, let's discuss the result. We care about the way the values of $l_1$ and $l_2$ varying with the variance of $k$ and $\theta$, for two reasons. First, one wave packet can be expanded in the basis of plane waves with different frequencies rang from $-\infty$ to $+\infty$. Second, in some situation in physics, one may consider the propagation of a wave packet, and in the route of the propagation there are some disturbance which is equal to the quantum walks introduced above. What's more, the disturbance is variable or even random, which equals to the variance or randomness of the value of $\theta$. So in Fig.\,1 we plot $L_1$ and $L_2$ acting as the function of $\theta$. We can see, both $L_1$ and $L_2$ are almost zero for some value of $\theta$, while in some other areas one of them would be much larger than the other. We choose $\theta=5.55$ to plot the relationship between $L_1$ $L_2$ and $k$ in Fig.\,2, where $L_2$ is much larger than $L_1$.

\subsection{No measurement at intermediate time steps}

  Now we improve the walks in Ref.\,\cite{Aharonov1} to the non-measurement case, which means we will not measure the state of the plane wave in the intermediate process, and let it evolute with the interference between the possible paths. This result would be more complicated.

  We suppose the initial state is
  \begin{equation}
  |\Psi(0)\rangle=(a_{\scriptscriptstyle R}|R\rangle+a_{\scriptscriptstyle L}|L\rangle)\otimes e^{ik(x-l_0)},
  \end{equation}

  the space operator is
  \begin{eqnarray}
  \hat{U} &=& \hat{S}\otimes\hat{P}_R+\hat{S}^\dagger\otimes\hat{P}_L \nonumber \\
  &=& |R\rangle\langle R|\otimes e^{-i\hat{P}l/\hbar}+|L\rangle\langle L|\otimes e^{i\hat{P}l/\hbar} \nonumber \\
  &=& \left(\begin{array}{cc}e^{-i\hat{P}l/\hbar} & 0  \\0 & e^{i\hat{P}l/\hbar} \end{array}\right) \ ,
  \end{eqnarray}

  and the evolution operator is
  \begin{eqnarray}
  \hat{E'}=\hat{R}(\theta)\hat{S} &=& \left(\begin{array}{cc}{\cos\theta} & {-\sin \theta}  \\{\sin\theta} & {\cos\theta} \end{array}\right) \left(\begin{array}{cc}e^{-i\hat{P}l/\hbar} & 0  \\0 & e^{i\hat{P}l/\hbar} \end{array}\right) \nonumber \\
&=& \left(\begin{array}{cc}e^{-i\hat{P}l/\hbar}\cos \theta & -e^{i\hat{P}l/\hbar}\sin \theta  \\e^{-i\hat{P}l/\hbar}\sin \theta & e^{i\hat{P}l/\hbar}\cos \theta \end{array}\right) \ .
  \end{eqnarray}

  The state after $t$ steps will be
  \begin{equation}
  |\Psi(t)\rangle=\hat{E'}^t|\Psi(0)\rangle=\left(\begin{array}{c} \varphi_{\scriptscriptstyle R}(t)\\ \varphi_{\scriptscriptstyle L}(t) \end{array}\right) \ .
  \end{equation}

  To solve this equation we can get the eigenstates and eigenvalues of matrix $\hat{M}$ first:
  $$ \hat{M}\left(\begin{array}{c} a \\ b_1 \end{array}\right)=\lambda_1 \left(\begin{array}{c} a \\ b_1 \end{array}\right) \ ,\ \ \ \ \ \hat{M}\left(\begin{array}{c} a \\ b_2 \end{array}\right)=\lambda_2 \left(\begin{array}{c} a \\ b_2 \end{array}\right) .$$
  $|\Psi(0)\rangle$ can be expanded in the two eigenstates:
  $$|\Psi(0)\rangle=\left(\begin{array}{c} a_{\scriptscriptstyle R}e^{ik(x-l_0)} \\ a_{\scriptscriptstyle L}e^{ik(x-l_0)}\end{array}\right)=m\left(\begin{array}{c} a \\ b_1 \end{array}\right)+n\left(\begin{array}{c} a \\ b_2 \end{array}\right),
$$
  so the final state is
  \begin{eqnarray*}
  |\Psi(t)\rangle &=& \hat{E'}^t|\Psi(0)\rangle =\left(\begin{array}{c} \varphi_{\scriptscriptstyle R}(t)\\ \varphi_{\scriptscriptstyle L}(t) \end{array}\right)= \varphi_{\scriptscriptstyle R}(t)|R\rangle+\varphi_{\scriptscriptstyle L}(t)|L\rangle \nonumber \\
  &=& \lambda_1^{t}\left(\begin{array}{c} a \\ b_1 \end{array}\right)m+\lambda_2^{t}\left(\begin{array}{c} a \\ b_2 \end{array}\right)n .
  \end{eqnarray*}

  Through much calculation we can get the analytical solutions:
  \begin{equation}
  \varphi_{\scriptscriptstyle R}(t)=\sqrt{P_1}e^{ik(x-l_0+L_1)},\ \ \varphi_{\scriptscriptstyle L}(t)=\sqrt{P_2}e^{ik(x-l_0+L_2)},
  \end{equation}
  where the value of $P_1$, $P_2$, $L_1$, and $L_2$ are determined by $\theta$, $k$ and the steps $t$. The details of the mathematical expressions and derivations of them are in the Appendix.

  The result shows that in the end of the walking the plane wave is still a plane wave, with a change in phase and no variance in frequency. We will use these results to study the quantum walks of general waves.

\section{The Quantum walk of general wave packets}
  As we know, every wave packet can be expanded in the form of plane waves (or equivalently cosine wave):
  \begin{equation}
  f(x,0)=\int^{+\infty}_{-\infty}\tilde{f}(k)e^{ikx}{\rm d}k,
  \end{equation}
  where $f(x,0)$ is the initial space state of the wave packet. As long as we know the evolution of the plane wave $e^{ikx}$, we can use the integral (25) to get the evolution result of wave $f(x)$, and the $\tilde{f}(k)$ can be treated as the weight of the plane wave $e^{ikx}$.
  \begin{equation}
  \tilde{f}(k)=\frac{1}{2\pi}\int^{+\infty}_{-\infty}f(x,0)e^{-ikx}{\rm d}x \ .
  \end{equation}

\subsection{Do measuring after each step}

  After one step, the plane wave $e^{ikx}$ develops into:
  \begin{equation}
  \sqrt{p_{\scriptscriptstyle R}}|R\rangle\otimes e^{ik(x+l_1)}+\sqrt{p_{\scriptscriptstyle L}}|L\rangle\otimes
  e^{ik(x+l_2)}.
  \end{equation}

  After $t$ steps, with measuring the coin state each step, the wave packet would develop into
  \begin{equation}
  e^{ik\left(x+nl_1+(t-n)l_2\right)}\ ,\ \ \ \ n\in [0,\ t],
  \end{equation}
  with the probability $p=\left(\begin{array}{c} t \\ n \end{array}\right)p^{n}_{\scriptscriptstyle R}p^{t-n}_{\scriptscriptstyle L}$. So one can study the quantum walks of any wave packet with the intermediate measurements, using Eq.\,(25).

\begin{figure}[tbph]
\centering
\includegraphics[width= 3in]{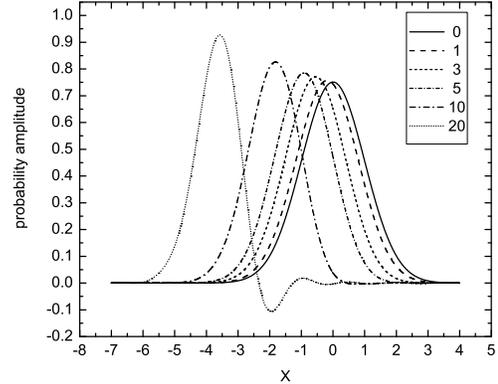}
\caption{The probability amplitude distribution for the Gaussian wave. with the same condition in the Ref.\,\cite{Aharonov1}, we choose the parameters $ \theta=-\arctan(0.9)$, $l=0.01$ to make the distance the wave moves each step much larger than $l$. This figure gives the shapes of the packet after 1, 3, 5, 10, and 20 steps, respectively.}\label{fig3}
\end{figure}
\begin{figure}[tbph]
\centering
\includegraphics[width= 3in]{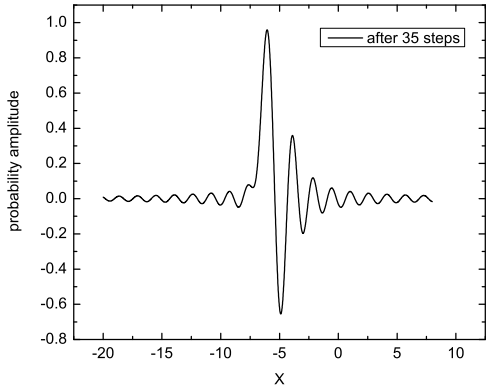}
\caption{The probability amplitude distribution for the Gaussian wave after 35 steps.}\label{fig4}
\end{figure}

  For a clear insight, we can give a typical example --- the walks of the Gaussian wave packet $f(x,0)={\rm exp}\left(-x^2/2\right)/\pi^{1/4}$. We consider the case that the result of the coin state measured each step would be $|L\rangle$. That is to say, the wave moves left each step. If the wave moves left each step, then the result after $t$ steps will be
  \begin{equation}
  f(x,0) \xrightarrow[]{}C\int^{+\infty}_{-\infty}\tilde{f}(k)\left(\sqrt{p_{\scriptscriptstyle L}}\right)^t|L\rangle\otimes e^{ik(x+tl_2)}{\rm d} k \ ,
  \end{equation}
  where $C$ is the normalization coefficient, to make sure the whole probability of the wave packet be unit; $p_{\scriptscriptstyle L}$ and $l_2$ are given by equations (17) and (19); and by Eq.\,(26), we have
\begin{equation}
\tilde{f}(k)=1/\sqrt{2\pi}\exp(-k^2/2).
\end{equation}

  Fig.\,3 and Fig.\,4 indicate that the wave will be split into several small waves after many steps of quantum walks. If $\theta$ is random and changing at each step, the results will be more interesting, here we will not proceed any more.

\subsection{No measurement at intermediate time steps}

  Assume that the initial state is
  $$(a_{\scriptscriptstyle R}|R\rangle+a_{\scriptscriptstyle L}|L\rangle)\otimes f(x,0),$$
  after $t$ steps it will evolute into
  \begin{equation}
  \int\limits^{+\infty}_{-\infty}\tilde{f}(k)\left(\sqrt{P_1}|R\rangle\otimes e^{ik(x+L_1)}+\sqrt{P_2}|L\rangle\otimes
  e^{ik(x+L_2)}\right){\rm d} k,
  \end{equation}
  where the parameters $P_1$, $L_1$, $P_2$, $L_2$ are functions of $k$ and $\theta$, which are given in the Appendix.
  \begin{figure}[tbph]
\centering
\includegraphics[width= 3in]{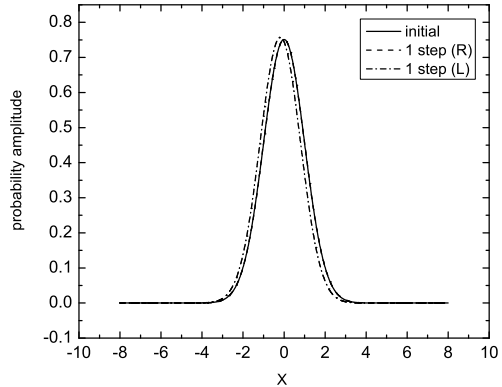}
\caption{The probability amplitude distribution for the Gaussian wave after one step. The result is the same to the former case. One wave moves left with much larger distance than $l$, and the other moves right with the distance similar to $l$ and so almost coincide with the initial Gaussian wave packet.}\label{fig5}
\end{figure}

\begin{figure}[tbph]
\centering
\includegraphics[width= 3in]{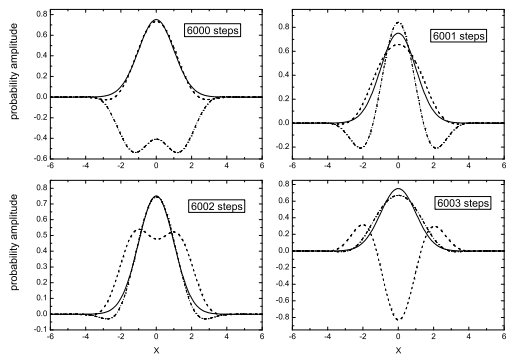}
\caption{The probability amplitude distribution for the Gaussian wave after 6000 to 6003 steps. The solid lines represent the initial Gauss wave; the dotted-dashed lines represent the waves with $|R\rangle$ coin state; the dashed lines represent the waves with $|L\rangle$ coin state.}\label{fig6}
\end{figure}

  We give the results by simulating the Gaussian wave: as same as the former case, we choose the parameters $\theta=-\arctan 0.9$, $l=0.01$, $a_{\scriptscriptstyle R}=a_{\scriptscriptstyle L}=1/\sqrt{2}$, $f(x,0)=\exp(-x^2/2)/\pi^{1/4}$. In figure 5 we note the fact that after one step the result is the same to the case of doing measurement at each step. We also find that when $t$ is not large enough the wave packet will not change much, and then we plotted the evolution of the Gaussian wave packet after $6000$ steps in Fig.\,6 to 9, the shapes of waves have much difference with initial shape, and would be split to some small waves. These figures indicate that when the wave has only one peak, the quantum walks will make little change on its shape at each step, but if a wave has two or more peaks the quantum walk will make large difference to its shape at each step for the reason that these peaks may interference each other during the walks.

\section{Conclusions}

  We have introduced a new idea --- the quantum walks of waves, and proposed a method to study and solve this problem. First, to simplify the calculation we studied the walks of plane wave, in which we considered two situations: with measurement in each step and without measurement in the intermediate process. Hence, we studied the quantum walks of general waves using the results of plane wave, which also includes the two situation mentioned above. Moreover, we have simulated some special wave packets. In the whole process, we find the behavior of waves quantum walking is much different from the counterparts of particles and also classical situation. By the way, the short-wave limit of wave is just delta function, which is equivalently particle.

  The quantum walks of waves can be used in many areas of physics, such as the propagation of light field in fibers, the transmission of signal with a width, and so on. In the propagation and transmission, there would be much disturbance such as reflection, refraction, and scattering, which are equal to the quantum walks with variable values of $\theta$. That is to say, the value of $\theta$ would be random in the process of quantum walking.

\begin{acknowledgments}
  We would like to thank Da Wei, Hao-Tian Wang, Guang-Wei Deng, and Ling-Lin Yu to give the support, some inspiring opinions and suggestions in the process of calculating and simulating. This work was funded by the National Fundamental Research Program, the National Natural Science Foundation of China (Grants No. 60621064 and No. 10974192), the Funds from the Chinese Academy of Sciences, and the K. C. Wong Foundation.
\end{acknowledgments}


\section*{APPENDIX: THE RESULT OF PLANE WAVE WITHOUT MEASUREMENT }
  To get the eigenvalues and eigenstates of $\hat{M}$, we notice that, for the plane wave $e^{ikx}$, the operator $e^{\pm i\hat{P}l/\hbar}$ is equivalent to $e^{\pm ikl}$.
$$
\hat{M}= \left(\begin{array}{cc}e^{-i\hat{P}l/\hbar}\cos \theta & -e^{i\hat{P}l/\hbar}\sin \theta  \\e^{-i\hat{P}l/\hbar}\sin \theta & e^{i\hat{P}l/\hbar}\cos \theta \end{array}\right)\ ,$$

$$ \hat{M}\left(\begin{array}{c} a \\ b_1 \end{array}\right)=\lambda_1 \left(\begin{array}{c} a \\ b_1 \end{array}\right) \ ,\ \ \ \  \ \hat{M}\left(\begin{array}{c} a \\ b_2 \end{array}\right)=\lambda_2 \left(\begin{array}{c} a \\ b_2 \end{array}\right). $$

  We can get:

  $$a=e^{i\hat{P}l/\hbar}\sin \theta,$$
  \begin{eqnarray*}
  b_{1,2}=&-&\frac{1}{2}(e^{i\hat{P}l/\hbar}-e^{-i\hat{P}l/\hbar})\cos \theta  \\
  &\pm&\frac{1}{2}\sqrt{(e^{i\hat{P}l/\hbar}-e^{-i\hat{P}l/\hbar})^2\cos ^2 \theta-4\sin ^2 \theta},
  \end{eqnarray*}
  \begin{eqnarray*}
  \lambda_{1,2}&=&\frac{1}{2}(e^{i\hat{P}l/\hbar}+e^{-i\hat{P}l/\hbar})\cos \theta  \\
  &\ &\mp\frac{1}{2}\sqrt{(e^{i\hat{P}l/\hbar}-e^{-i\hat{P}l/\hbar})^2\cos ^2 \theta-4\sin ^2 \theta},
  \end{eqnarray*}
  $$\lambda_1=-ie^{i\alpha}\ ,\ \ \ \ \ \ \lambda_2=ie^{-i\alpha},$$

  with
  \begin{equation}
  \alpha=\arctan \frac{\cos kl\cos \theta}{\sqrt{\sin ^2 kl \cos^2 \theta + \sin ^2 \theta}}.
  \end{equation}
  After $t$ steps, the wave develops into
  $$ \varphi_{\scriptscriptstyle R}(t)=\sqrt{P_1}e^{ik(x-l_0+L_1)}\ \ {\rm and}\ \ \varphi_{\scriptscriptstyle L}(t)=\sqrt{P_2}e^{ik(x-l_0+L_2)}$$
  If $t$ is even\ \ \ \ $P_1=A^2+B^2,\ \ \ P_2=C^2+D^2,$
  $$\tan(kL_1-t\pi/2)=B/A,\ \ \tan(kL_2-t\pi/2)=D/C.$$
  If $t$ is odd\ \ \ \ \ $P_1=E^2+F^2,\ \ \ P_2=G^2+H^2,$
  $$\tan(kL_1-t\pi/2)=F/E,\ \ \tan(kL_2-t\pi/2)=H/G.$$
  with
  $$A=a_{\scriptscriptstyle R}\cos t\alpha +a_{\scriptscriptstyle L}\cos kl\sin \theta\sin t\alpha/Q,$$
  $$B=(a_{\scriptscriptstyle R}\cos \theta+a_{\scriptscriptstyle L}\sin \theta)\sin kl\sin t\alpha/Q,$$
  $$C=a_{\scriptscriptstyle L}\cos t\alpha -a_{\scriptscriptstyle R}\cos kl\sin \theta\sin t\alpha/Q,$$
  $$D=(a_{\scriptscriptstyle R}\sin \theta-a_{\scriptscriptstyle L}\cos \theta)\sin kl\sin t\alpha/Q,$$
  $$E=-(a_{\scriptscriptstyle R}\cos \theta+a_{\scriptscriptstyle L}\sin \theta)\sin kl\cos t\alpha/Q,$$
  $$F=-a_{\scriptscriptstyle R}\sin t\alpha +a_{\scriptscriptstyle L}\cos kl\sin \theta\cos t\alpha/Q,$$
  $$G=(-a_{\scriptscriptstyle R}\sin \theta+a_{\scriptscriptstyle L}\cos \theta)\sin kl\cos t\alpha/Q,$$
  $$H=-a_{\scriptscriptstyle L}\sin t\alpha -a_{\scriptscriptstyle R}\cos kl\sin \theta\cos t\alpha/Q.$$

  Here
  $$Q=\sqrt{\sin^2kl\cos^2\theta+\sin^2\theta}.$$



\end{document}